%% file: top.tex
\documentclass[conference]{IEEEtran}
\usepackage{amsmath,amssymb,amsfonts}
\usepackage{graphicx}
\usepackage{textcomp}
\usepackage{xcolor}
\usepackage{algorithm}
\usepackage{algpseudocode}
\usepackage[hidelinks]{hyperref}

\usepackage{array,tabularx}
\usepackage{siunitx}
\usepackage{needspace} 
\newcolumntype{Y}{>{\centering\arraybackslash}X} 

\def\BibTeX{{\rm B\kern-.05em{\sc i\kern-.025em b}\kern-.08em
    T\kern-.1667em\lower.7ex\hbox{E}\kern-.125emX}}

\usepackage{cite}

\newcommand{\afterfigure}{\par\vspace{-\parskip}\noindent\ignorespaces}

\begin{document}

\title{A Scalable NorthPole System with \\ End-to-End Vertical Integration for \\ Low-Latency and Energy-Efficient LLM Inference\\}

\author{\IEEEauthorblockN{
Michael V. DeBole,
Rathinakumar Appuswamy, 
Neil McGlohon,
Brian Taba, 
Steven K. Esser, \\ 
Filipp Akopyan, 
John V. Arthur, 
Arnon Amir, 
Alexander Andreopoulos, 
Peter J. Carlson, \\
Andrew S. Cassidy, 
Pallab Datta, 
Myron D. Flickner, 
Rajamohan Gandhasri, 
Guillaume J. Garreau, \\ 
Megumi Ito, 
Jennifer L. Klamo, 
Jeffrey A. Kusnitz, 
Nathaniel J. McClatchey, 
Jeffrey L. McKinstry, \\
Tapan K. Nayak, 
Carlos Ortega Otero, 
Hartmut Penner,
William P. Risk,
Jun Sawada, \\
Jay Sivagnaname,  
Daniel F. Smith, 
Rafael Sousa, 
Ignacio Terrizzano, 
Takanori Ueda, \\
Trent Gray-Donald$^\dagger$, 
David Cox, 
Dharmendra S. Modha* 
}
\IEEEauthorblockA{\textit{IBM Research, $^\dagger$IBM Software}, 
*dmodha@us.ibm.com}}

\maketitle

\setcounter{topnumber}{3}
\setcounter{bottomnumber}{3}
\setcounter{totalnumber}{6}
\setcounter{dbltopnumber}{2}

\renewcommand{\topfraction}{0.9}
\renewcommand{\bottomfraction}{0.8}
\renewcommand{\textfraction}{0.05}
\renewcommand{\floatpagefraction}{0.8}

\renewcommand{\dbltopfraction}{0.9}
\renewcommand{\dblfloatpagefraction}{0.8}

\setlength{\textfloatsep}{8pt plus 2pt minus 2pt}
\setlength{\dbltextfloatsep}{8pt plus 2pt minus 2pt}
\setlength{\floatsep}{6pt plus 2pt minus 2pt}

\input{content}

\bibliographystyle{IEEEtran.bst}
\bibliography{references}

\vspace{12pt}

\end{document}

%% file: content.tex
\begin{abstract}

A vertically integrated, end-to-end, research prototype system combines 288 NorthPole neural inference accelerator cards, offline training algorithms, a high-performance runtime stack, and a containerized inference pipeline to deliver a scalable and efficient cloud inference service. The system delivers 115~peta-ops at 4-bit integer precision and 3.7~PB/s of memory bandwidth across 18~2U servers, while consuming only 30~kW of power and weighing 730~kg in a 0.67~m\textsuperscript{2} 42U rack footprint. The system can run 3 simultaneous instances of the 8-billion-parameter open-source IBM Granite-3.3-8b-instruct model at 2,048 context length with 28 simultaneous users and a per-user inter-token latency of 2.8 ms. The system is scalable, modular, and reconfigurable, supporting various model sizes and context lengths, and is ideal for deploying agentic workflows for enterprise AI applications in existing data center (cloud, on-prem) environments. For example, the system can support 18 instances of a 3-billion-parameter model or a single instance of a 70-billion-parameter model.
\end{abstract}


\section{Introduction}

Large language models have become a pervasive form of computing, and while the current paradigm has been to push frontier models for all applications, it is becoming evident that “Faith in God-like large language models is waning” \cite{economist-2025-llm-waning}. In fact, by continuing along this trajectory, global energy requirements for AI-focused data centers are projected to reach double-digit percentages of total electricity consumption by 2030, with individual facilities requiring up to 1 gigawatt or more of dedicated power---driving both infrastructure and cooling costs toward potentially unsustainable or unprofitable levels \cite{Noffsinger2025}\cite{Pilz2025}.

However, for many business applications, frontier models containing trillions of parameters may prove less useful and cost efficient than much smaller language models with only a tenth or even a hundredth as many parameters\cite{wsg-2025-small-models}. These purpose-built Large Language Models (LLMs) not only improve task performance on a focused domain but also offer the opportunity to apply emerging new computing architectures, for example, IBM’s NorthPole, to deliver low latency and high energy efficiency, making the use of AI a sustainable and profitable component of a business.
\Needspace{12\baselineskip} 
\begin{figure}[!tp]
  \centering
  \includegraphics[width=\columnwidth,keepaspectratio]{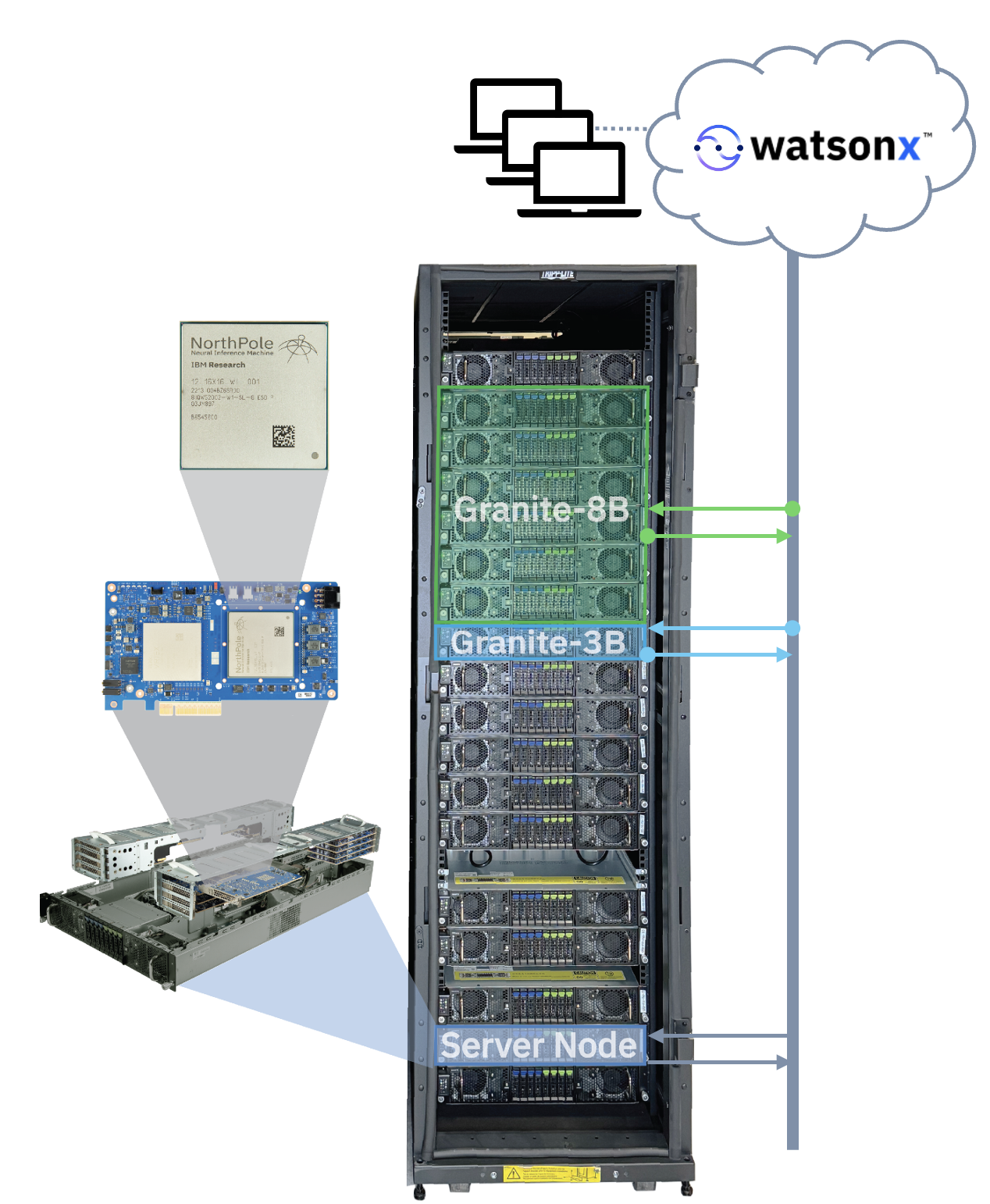} 
  \caption{End-to-end view of the NorthPole LLM inference system. The NorthPole chip (upper left) is a highly parallel neural inference accelerator deployed in a PCIe form factor by a NorthPole card (middle left), 16 of which are hosted by a 2U server (lower left) to form a NorthPole LLM server node (lower right), up to 18 of which make up a NorthPole LLM inference rack. Each rack can run multiple LLM instances simultaneously at high throughput and low latency. Larger models can be deployed by connecting multiple racks. Access to the NorthPole-accelerated models are provided via the IBM watsonx web interface and API.
}
  \label{fig:288Rack}
\end{figure}
\afterfigure NorthPole is an AI inference chip designed to push the frontiers of time, space, and energy \cite{np_science}, currently available as a research prototype in a PCIe card form factor. First demonstrated in \cite{np_hpec_llm} as a single LLM server node with 16 NorthPole PCIe cards, the {\bf first contribution} of this paper is to extend that work by combining 18~2U~servers in a rack, resulting in a 288 NorthPole LLM inference system. The overall system, described in Section~\ref{sec:hardware}, delivers 115~peta-ops at 4-bit integer precision, has 3.7 PB/s of memory bandwidth, consumes 30~kW of power, weighs 730~kg, and fits within a 0.67~m$^2$ 42U footprint. It is noteworthy that the system does not require specialized communication fabrics, custom hardware integration, liquid cooling, or facility power upgrades, making it easy to deploy at scale in existing cloud or on-prem data center environments.

As a motivating example, this paper focuses on the 8-billion-parameter open-source IBM Granite-3.3-8b-instruct model \cite{granite-3.3-8b-instruct}, although the results are extensible to, for example, recent 20-billion and 120-billion-parameter models from OpenAI. These small-to-medium-size models, with small-to-medium context lengths, are increasingly seen as the future for many enterprise AI applications \cite{belcak2025smalllanguagemodelsfuture}\!\cite{pareja2024unveilingsecretrecipeguide}. To map the model to the NorthPole system, a pipelined parallelism strategy is employed. Model parameters and KV cache reside entirely in on-chip memory, which reduces communication between cards to the point that they can be interconnected within each server by PCIe and between servers by low-latency 200~GbE NICs. This mapping strategy, which is the {\bf second contribution} of this paper, is described in Section~\ref{sec:pipeline_parallelism}. 

Section~\ref{sec:cloud_svc} and Section~\ref{softwareruntimestack} describe the {\bf third contribution} of this paper, namely, a containerized inference pipeline and a high-performance runtime stack to deliver a scalable and efficient cloud inference service. Moreover,  \autoref{fig:watsonx} demonstrates the direct integration of the cloud-based inference service with IBM watsonx Orchestrate \cite{IBMwatsonxOrchestrate}, enabling a variety of applications which invoke watsonx APIs to leverage the NorthPole system seamlessly.

Finally, the {\bf fourth contribution} of this paper, described in Section~\ref{sec:results}, shows how it is possible to fine-tune a model quantized to 4-bit precision to match the accuracy of a model originally trained at bfloat16 precision, using a recently developed quantization-aware training algorithm called SiLQ\cite{esser2025silqsimplelargelanguage} 

Combining all four contributions together results in an end-to-end, vertically integrated, NorthPole inference system that can run 3 simultaneous instances of the IBM Granite-3.3-8b-instruct model at 2,048 context length with 28 simultaneous users at a per-user inter-token latency of 2.8~ms. The same system can run 18 instances of a 3-billion-parameter model at the same context length and number of users, achieving an inter-token latency of 1~ms \cite{np_hpec_llm}.

\section{Hardware}\label{sec:hardware}

The hardware building blocks of the NorthPole LLM inference system comprise the NorthPole chip, PCIe card, LLM server node, and LLM inference rack, as shown in \autoref{fig:288Rack}. The chip and card are generic to any neural inference application, while the server node and inference rack designs are based on the power and bandwidth requirements of an LLM application.

\subsection{NorthPole chip}
The NorthPole chip is the central component of the inferencing hardware system, comprising 22~billion transistors fabricated in 800~mm\textsuperscript{2} with the GlobalFoundries 12-nm process. It embodies a custom neural inference architecture derived from a set of design axioms that collectively yield exceptional performance on time, space, and energy metrics\cite{np_science}\!\cite{np_isscc}.

The chip contains a 16\ensuremath{\times}16~array of cores, each with its own compute, communication, control, and memory resources, promoting a distributed view of computation. It has native support for 8/4/2-bit integer and 16-bit floating-point precisions and contains 224~MB of on-chip memory, consisting of 192~MB spread across the core array to store weights and intermediate tensors and a 32~MB framebuffer to stage input and output tensors for off-chip communication.

To maximize data locality and minimize energy consumption, all weights and intermediate activations reside entirely within on-chip memory at inference time, ensuring that most data movement is local within individual cores and effectively eliminating the need for costly transfers to and from off-chip memory during neural network execution. In sharp contrast to conventional architectures like GPUs, whose memory models require frequent transfer of weights and intermediate computations between on- and off-chip memory during each inference, NorthPole only transfers input and output tensors off-chip, drastically reducing the external bandwidth requirements and energy costs associated with data movement.
 
\subsection{NorthPole card}
The NorthPole card deploys the NorthPole chip in a single-wide PCIe form factor. In addition to components for standard board functions like power sequencing and clock generation, the card uses an FPGA to implement the PCIe endpoint logic, DMA engines, and datapath control required to transfer data between host memory and the NorthPole chip. The FPGA also implements logic for cards to exchange tensors directly with each other via the PCIe fabric, eliminating the need for costly memory copies to and from host memory when passing output tensors between cards. The low latency of direct card-to-card communication is a critical feature for mapping transformer-based models using pipeline parallelism.

For LLMs and other transformer-based workloads, the NorthPole card typically consumes less than 55~W, allowing air cooling using discrete heat sinks for the chip and FPGA and appropriate airflow management within the server.

\begin{figure}[tp]
  \centering
  \includegraphics[width=\linewidth]{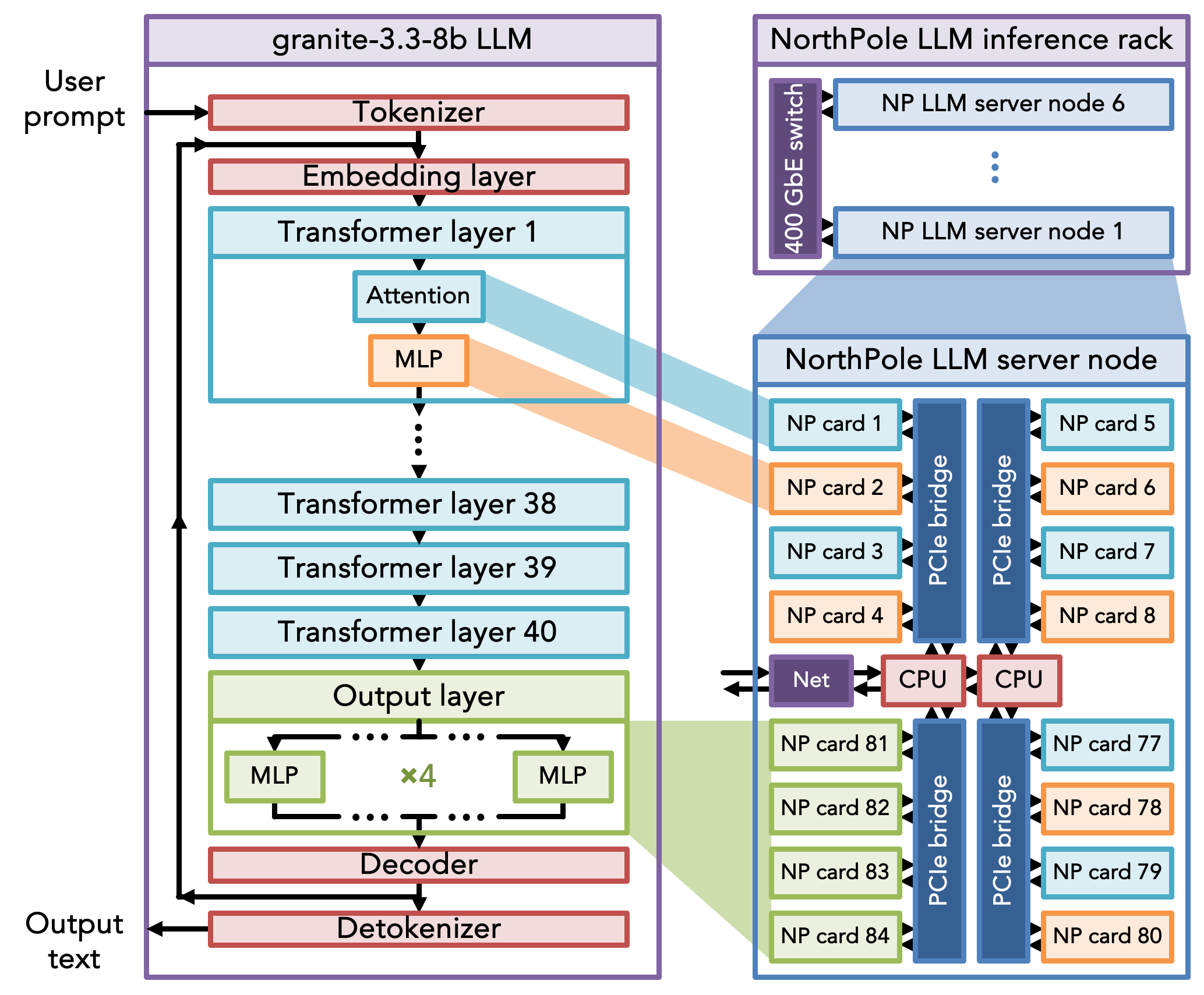} 
  \caption{Mapping an 8-billion-parameter LLM to NorthPole. The attention and multi-layer perceptron (MLP) blocks of each of the 40 transformer layers of the Granite-3.3-8b-instruct model (left) are mapped to separate NorthPole cards (lower right) using pipeline parallelism. The output layer is split across 4 NorthPole cards using tensor parallelism. The full model uses 84 cards in 6 NorthPole LLM server nodes that are interconnected via 200~GbE and occupy 12U of a NorthPole LLM inference rack (upper right).
}
  \label{fig:granite-8b-mapping}
\end{figure}

\subsection{NorthPole LLM Server Node}
Server selection is greatly facilitated by the low power consumption of NorthPole and specifically for LLM networks which are mapped to multiple cards and exploit pipeline parallelism. As no specialized cooling solutions are required, an off-the-shelf server can be chosen to maximize PCIe card density while minimizing physical footprint within a rack. The server requires a motherboard architecture that provides the necessary PCIe bandwidth to each card, as well as sufficient compute resources (CPUs and cores) and DDR memory capacity and bandwidth to meet application-specific latency and throughput targets, since the host processor is responsible for non-neural operations like tokenization.

To meet these goals, the NorthPole LLM server node configures a Gigabyte G292-2G0 server\cite{gigabyte_g292_2g0} with dual 3rd-generation Intel Xeon Gold 6354 Processors (18~cores each, 39~MB of L3~cache), 1~TB of DDR4 Memory, and Mellanox CX-6 adapters capable of 200~GbE and RoCE for low-latency, high-throughput, inter-system communication. These servers offer the highest commercially available PCIe card density within a single integrated chassis, accommodating up to 16 NorthPole cards in 2U of rack space.

When fully populated with 16~cards, a standalone NorthPole LLM server node running a 3-billion-parameter variant of the open-source IBM Granite-8b-code-base-4k model\cite{granite-8b-code-base-4k} with 4-bit weights and activations can deliver almost the same accuracy as the full-precision model at a massive 28,356~token/s of system throughput and sub-1~ms/token per-user latency while, consuming only 672~W of aggregated card power\cite{np_hpec_llm}.

\subsection{NorthPole LLM Inference Rack}
Since data center requirements are not one-size-fits-all, the NorthPole LLM inference rack aims to demonstrate one possible configuration suited for deployments constrained to a maximum rack power budget of 40~kW and air-based cooling. Within those constraints, a rack was configured with 18 NorthPole LLM server nodes. A separate rack houses a 32-port 400~GbE Mellanox switch enabling all-to-all connectivity between nodes within each rack. 

Each NorthPole LLM inference rack contains 288 NorthPole cards, providing 60, 115, and 230~peta-ops per second for 8-bit, 4-bit, and 2-bit integer operations, respectively, adhering to the 40~kW envelope suitable for entirely air-based cooling. A fully populated rack weighs approximately 730~kg in a 0.67~m\textsuperscript{2} 42U footprint (1090~kg/m\textsuperscript{2}, 45~kW/m\textsuperscript{2}).

\section{Mapping LLMs to NorthPole}\label{sec:pipeline_parallelism}

Mapping an LLM to NorthPole involves three choices: how to partition the model across multiple cards, how to quantize weights and activations to fit more parameters and larger KV caches into on-chip memory, and how to choose mini-batch and micro-batch sizes for efficient pipeline utilization.

\begin{figure}[tp]
  \centering
  \includegraphics[width=\linewidth]{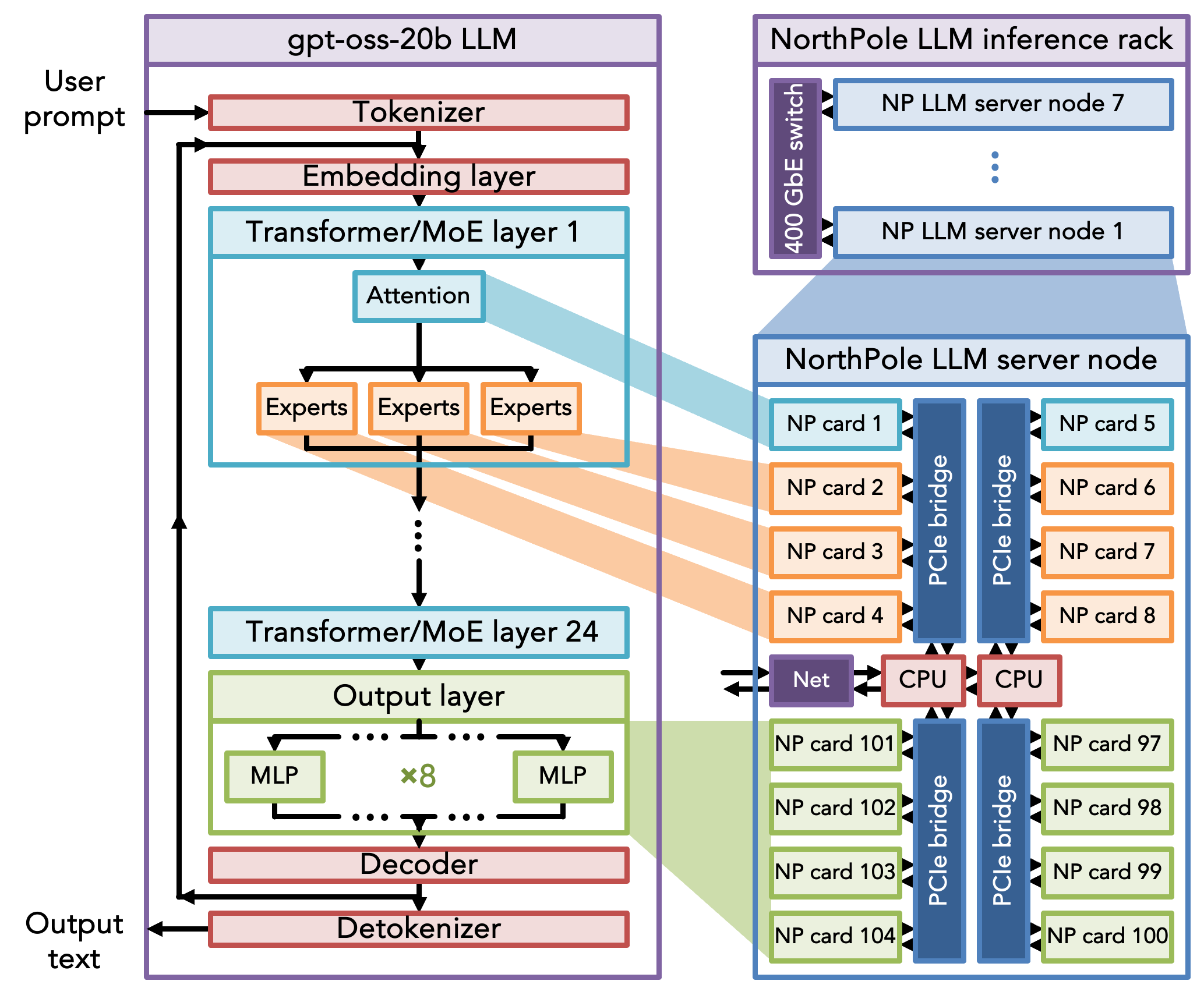} 
  \caption{Mapping a 20-billion-parameter LLM to NorthPole. The attention and expert blocks of each of the 24 transformer/MoE layers of the gpt-oss-20b model are mapped to separate NorthPole cards using tensor and pipeline parallelism. The full model uses 104 cards in 7 NorthPole LLM server nodes. The 120-billion-parameter gpt-oss-120b model (not shown) can likewise be mapped using 11 cards for experts in each of the 36 layers, for a total of 440 cards in 28 server nodes across 2 inference racks.
}
  \label{fig:gpt-oss-20b-mapping}
\end{figure}

\subsection{Model Partitioning}

The strategy for mapping a model to NorthPole cards is inspired by three key observations. First, once a single transformer layer has been mapped to one or more cards, multiple such layers can be stacked using pipeline parallelism all the way up to the output layer. Second, if the weights and KV cache for all LLM layers reside entirely on-chip, only the small embedding tensor needs to be communicated between layers at inference time, which is well within the bandwidth of PCIe Gen3\ensuremath{\times}8. Third, a NorthPole system with that pipeline-parallel bandwidth can readily be assembled by populating an off-the-shelf server with 16 NorthPole cards, and larger models can be composed by networking multiple servers together.

These observations suggest a strategy that partitions each transformer layer across its own set of NorthPole cards using a combination of data, tensor, attention head, and pipeline parallelism strategies, and exclusively uses pipeline parallelism between layers in the style of GPipe \cite{huang2019gpipeefficienttraininggiant}. The output layer is split using tensor parallelism across one or more NorthPole cards, depending on vocabulary size \cite{shazeer2018meshtensorflowdeeplearningsupercomputers}\!\cite{shoeybi2020megatronlmtrainingmultibillionparameter}.

\subsection{Quantization}

NorthPole represents weights and activations at 8-bit, 4-bit, or 2-bit integer precisions. Every layer in the quantized model can set different precisions for its activations, KV cache, and weights. Lower precisions allow more parameters and larger caches to reside within on-chip memory, reducing the number of cards needed to map each layer, but higher precisions make it easier for quantization-aware training recipes like SiLQ to achieve the accuracy of the unquantized model by fine-tuning on a tiny fraction of the original training data\cite{esser2025silqsimplelargelanguage}.

For example, using 8-bit activations, 8-bit caches, and 4-bit weights (A8-C8-W4) in all layers, the attention and multi-layer perceptron (MLP) blocks in each of the 40 transformer layers in the IBM Granite-3.3-8b-instruct model\cite{granite-3.3-8b-instruct} can be mapped to 2 cards by pipeline parallelism, and the output layer to 4 cards by tensor parallelism, for a total of 84 cards, as shown in \autoref{fig:granite-8b-mapping}. Although outside the scope of this paper, a similar strategy can be used to map mixture-of-expert (MoE) models like OpenAI's gpt-oss-20b \cite{openai2025gptoss120bgptoss20bmodel}, as shown in \autoref{fig:gpt-oss-20b-mapping}. Table \ref{tab:model-configs} summarizes the NorthPole hardware resources required to map different models.

\begin{table}[!t]
\centering
\renewcommand{\tabularxcolumn}[1]{m{#1}}
\caption{Model configurations and corresponding hardware resources. The Granite-3.1 model uses 4-bit activations, caches, and weights (A4-C4-W4), and the other models use 8-bit activations and caches and 4-bit weights (A8-C8-W4).}
\begin{tabularx}{\linewidth}{|Y|Y|Y|Y|Y|}
\hline
\textbf{Model family} & \textbf{Parameters} & \textbf{NorthPole cards} & \textbf{Server nodes} & \textbf{Inference racks} \\ \hline
Granite-3.1 & 3B & 16 & 1 & 1 \\ \hline
Granite-3.3 & 8B & 84 & 6 & 1 \\ \hline
gpt-oss & 20B & 104 & 7 & 1 \\ \hline
gpt-oss & 120B & 440 & 28 & 2 \\ \hline
\end{tabularx}
\label{tab:model-configs}
\end{table}

\subsection{Pipeline Utilization}

During inference, the system processes a mini-batch of $N$ simultaneous user requests, divided into $M$ micro-batches that are pipelined through the NorthPole cards. Efficient hardware utilization favors large mini-batches that keep every card occupied with its own micro-batch, preventing pipeline bubbles, but larger mini-batches take longer to compute, increasing per-token latency. As a result, while pipeline parallelism has been exploited for LLM training, which has no latency constraints, its use for inference has been limited by the latency of the large mini-batch needed for efficient GPU utilization. For example, it was found that efficient training required the number of micro-batches to be about four times the number of GPU pipeline stages\cite{huang2019gpipeefficienttraininggiant}.

For NorthPole, the low-latency compute within the core array and 13~TB\,/\,s of on-chip memory bandwidth per chip give the system extremely low per-token latency, so the limiting factor in choosing $N$ is the on-chip memory available to store the KV cache for the  entire mini-batch. Constraining the KV cache to fit entirely on-chip also imposes a tradeoff between context length $L$ and the number of simultaneous users $N$.

In this paper, mini-batches were chosen that facilitate placing the entire KV cache within on-chip memory, and were divided into micro-batches of size 1 when the number of pipeline stages was 16 or more, or into larger micro-batches for smaller models with fewer pipeline stages. It was observed that a number of micro-batches equal to the number of NorthPole pipeline stages sufficed to keep pipeline idle time negligible. The ability of the NorthPole architecture to compute efficiently at such a small micro-batch size is key to achieving the performances reported in Section~\ref{sec:results}.

\begin{figure}[tp]
    \centering
    \includegraphics[width=\linewidth]{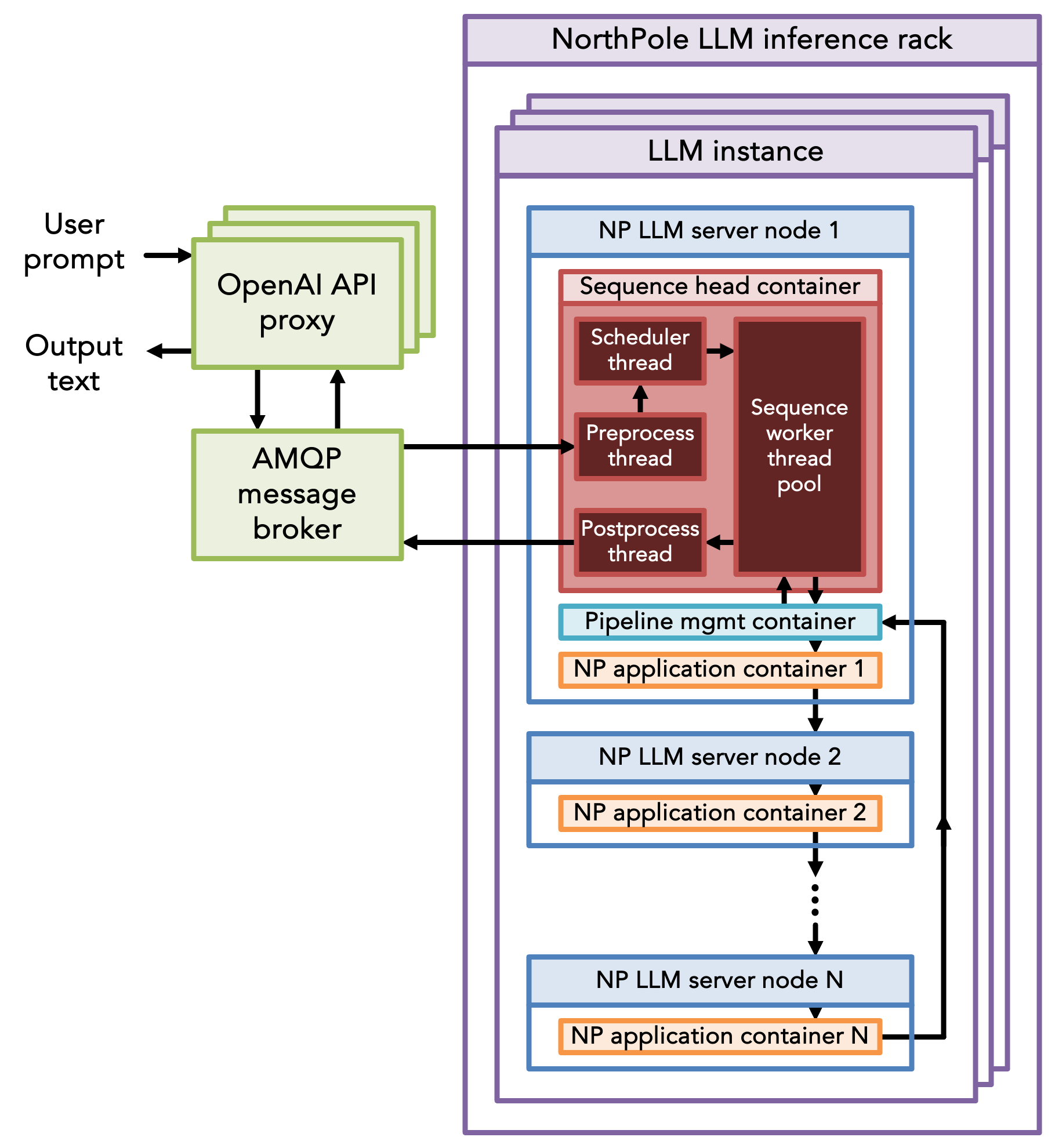}
    \caption{NorthPole LLM inference service. Each LLM instance runs on its own pipeline-parallel chain of one or more NorthPole LLM server nodes. Each server node hosts a NorthPole application container that controls, configures, and communicates with its NorthPole cards. The first server node in the chain hosts two additional containers: a pipeline management container to handle input and output for the server node pipeline, and a sequence head container for pre- and postprocessing tasks like tokenization and interacting with the cloud services (left) that connect the user with the NorthPole LLM inference system (right).}
    \label{fig:np-app-llm-architecture}
\end{figure}

\section{Cloud Inference Service}\label{sec:cloud_svc}

Once mapped to NorthPole hardware, the LLM model is instantiated in software as a containerized application deployed on a NorthPole LLM inference system and accessed through the cloud, as shown in \autoref{fig:np-app-llm-architecture}. Each LLM instance comprises a set of containers that collectively configure a chain of LLM server nodes to run a single model that can simultaneously generate independent sequences of tokens for a number of users that depends on model size and context length.

Front-end applications and other services interact with the system through a set of endpoints that implement OpenAI's streaming chat completions protocol\cite{openai_api}. Upon receiving a user prompt, the API endpoint component posts an inference task specifying the requested LLM model and service priority to the appropriate queue in an Advanced Message Queuing Protocol (AMQP) message broker, in this case RabbitMQ deployed in IBM Cloud. Each LLM instance subscribes to the task queue for the model it is running, and consumes a task from the queue whenever it is ready to process a new sequence. An LLM instance can subscribe to some or all priority levels for its model, making it easy to provide load balancing and uniform quality of service for users with different service-level entitlements.

Within the NorthPole LLM inference system, each LLM instance is composed of three types of containers. Every LLM server node runs its own NorthPole application container to configure, control, and communicate with the NorthPole cards it hosts, using the software runtime stack described in Section~\ref{softwareruntimestack}. The first LLM server node in the chain also runs a pipeline management container to coordinate the pipeline of NorthPole application containers, and a sequence head container to handle the external interface to the cloud services, perform non-neural computations like tokenization, and provide a pool of worker threads to manage the generation of individual token sequences.

\begin{enumerate}
    \item \textit{Sequence Head Container:} The sequence head container maintains a pool of sequence worker threads, one for every simultaneous user prompt. When sequence worker threads become available, the sequence head container pulls new user prompts from the subscribed AMQP queue and tokenizes them using a dedicated preprocessing thread. A scheduler thread dispatches each tokenized prompt to an available sequence worker in the pool. This enables the system to implement dynamic batching where multiple user queries can start and complete asynchronously relative to one another.
    \item \textit{Pipeline Management Container:} At startup, all NorthPole application containers configure their cards in parallel. The pipeline management container uses a ring-based consensus protocol to determine when all application containers have finished configuring their cards, then acts as a passthrough interface to send input to the first application container and receive output from the last application container.
    \item \textit{NorthPole Application Container:} Every LLM server node has its own NorthPole application container to configure each hosted card with its portion of the model. At inference time, the application container uses sockets to receive tensors generated by layers in upstream server nodes, and calls the runtime library to send the input tensors to the first card. Each card's output is relayed to the next card using direct card-to-card communication, with only the last card's output returned to the host for transmission over sockets to downstream server nodes.
\end{enumerate}

Each sequence worker manages the full life cycle of token generation for an individual sequence. During prompt prefill and iterative token generation, the sequence worker passes tokens via the pipeline management container to the first application container in the chain. Each application container executes the network layers allocated to its LLM server node and forwards their output tensors to downstream application containers. The final application container returns its output to the pipeline management container, which sends the generated token to the sequence worker. If the sequence worker determines that the stopping criteria have been met, it hands off the generated sequence to the postprocessor thread and returns to the thread pool. The postprocessor collects sequence statistics and sends the completed response back to the API endpoint component via the AMQP message broker’s response channel.

\section{Software Runtime Stack}\label{softwareruntimestack}
To interact with the NorthPole cards, an LLM application container interacts with a software runtime stack comprising two primary components: a user-space Linux driver for low-level hardware control and a runtime library that wraps the driver with a high-level API for host applications to load models onto each chip, send input tensors, and receive output tensors. NorthPole cards hosted by the same server node can communicate with each other directly over PCIe, without host intervention.

\subsection{User-Space Driver}
The user-space driver is not directly accessible to applications. It handles low-level hardware initialization and provides an abstracted interface for the runtime library to perform memory-mapped I/O (MMIO) and direct memory access (DMA) operations to and from the NorthPole chip on the NorthPole card. The driver allocates memory-mapped buffers as requested by the runtime library and configures DMA descriptors that instruct the DMA engines on the FPGA how to transfer data between host memory and card.

\subsection{Runtime Library}
The runtime library exposes a simple C++ API for host applications to interact with NorthPole using methods to load ELF-formatted model binaries, send input tensors asynchronously, and receive output tensors through a callback mechanism. The runtime library is multithreaded, so model loading, input submission, and output handling all happen concurrently while maintaining the required data dependency and ordering guarantees.

A critical responsibility of the runtime library is memory management of each chip's framebuffer, which serves as a staging area for future input tensors and previous output tensors so off-chip communication can be pipelined with on-chip computation of current tensors by the core array. To avoid overwriting input tensors before they are sent to the core array, or output tensors before they are sent off-chip, the runtime library provides methods to ensure that input tensors are only transferred to a card when enough space is available in its chip's framebuffer, and that tensor data is correctly placed within the framebuffer’s address space for model execution.

\subsection{Direct Card-to-Card Communication}
The NorthPole runtime stack provides native support for direct card-to-card (C2C) communication of tensors over PCIe between NorthPole cards that are hosted by the same server, without intervening card-to-host (C2H) and host-to-card (H2C) memory copies. To do this, the user-space driver implements mechanisms that enable direct C2C DMA transfers via IOMMU-mapped IOVA spaces, and the runtime library leverages three features implemented by the FPGA on the NorthPole card: output-input packet conversion, framebuffer credit tracking, and locally stored DMA descriptor chains.

\begin{enumerate}
    \item \textit{Output-Input Packet Conversion:} Output data packets from one NorthPole card are automatically converted into input data packets for the destination card. The conversion process also determines the correct placement of incoming data within the destination framebuffer for subsequent computation.
    \item \textit{Framebuffer Credits:} To manage framebuffer memory without host intervention, the FPGA maintains programmable credit counters that track the number of free framebuffer locations in its destination card(s). Sending an output tensor to occupy a framebuffer location in a destination card decrements the source card's credit counter for that framebuffer. If a credit counter reaches zero, further outputs are held at the source card until there is space at the destination. When the destination card has consumed a tensor as input, as indicated by transferring the resulting output tensor(s) to a downstream card or host, it frees its framebuffer location by sending a credit packet back to its source card to increment that framebuffer's credit counter.
    \item \textit{Locally Stored DMA Descriptor Chains:} To enable autonomous data transfers at inference time, DMA descriptor chains for all output and credit packets are precomputed by the runtime library during application initialization and stored locally in the corresponding FPGAs, substantially reducing host CPU overhead.
\end{enumerate}

At initialization, the runtime library configures the on-chip memories of all cards hosted by the server node with their portions of the model, and loads their FPGAs with precomputed DMA descriptor chains that define one or more virtual circuits of cards through which tensors can be routed when executing the model. The runtime library seamlessly toggles between virtual circuits, allowing the host application to run, for example, a MoE model using different subsets of experts for each execution without reconfiguring on-chip memories.

At inference time, the host application supplies the runtime library with the input tensors to be processed, along with a circuit identifier to specify the execution path. The runtime library transfers the input tensors to the designated entry cards for the selected circuit and configures output DMA descriptors on all participating cards, which autonomously exchange intermediate outputs and framebuffer credits using direct card-to-card communication that is transparent to the host application. The designated output cards transfer the final computations back to host memory, where they are received by the host application through asynchronous callback functions that it registered with the runtime library at initialization.

\begin{figure}[tp]
    \centering
    \includegraphics[width=.9\linewidth]{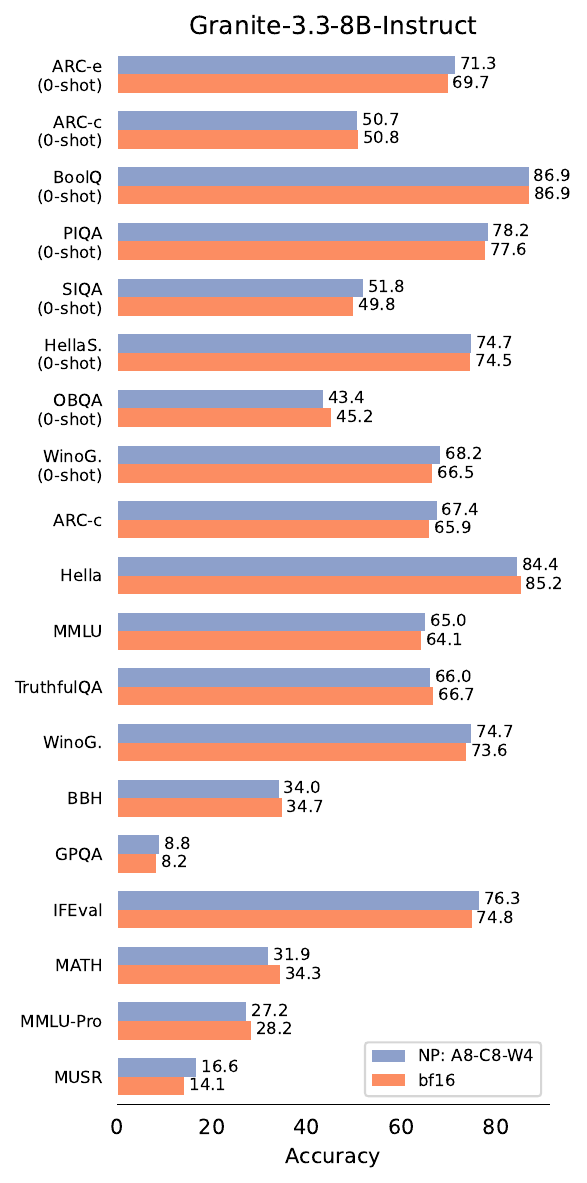}
    \caption{Accuracy of quantized (A8-C8-W4) Granite-3.3-8b-instruct model when run on NorthPole, compared to original bfloat16 (bf16) model, on 19 benchmarks, including common sense reasoning tasks, and tasks from versions 1 and 2 of the Open LLM Leaderboard.}
    \label{fig:accuracy}
\end{figure}

\begin{table*}[!t]
\centering
\renewcommand{\tabularxcolumn}[1]{m{#1}}
\caption[Latency and Comparison]{Latency and throughput measurements for Granite-3.3-8b-instruct within a single LLM instance \footnotemark.}
\begin{tabularx}{\textwidth}{|Y|Y|Y|Y|Y|Y|Y|}
\hline
\textbf{Context Length} & \textbf{Batch Size} & \textbf{$\mathbf{TTFT}_s$ (ms)} & \textbf{$\mathbf{ITL}_s$ (ms)} & \textbf{$\mathbf{ITPS}_B$} & \textbf{$\mathbf{OTPS}_B$} & \textbf{$\mathbf{EOTPS}_B$} \\ \hline
2k & 28 & 64.8 & 2.8 & 78996 & 10341 & 9552 \\ \hline
4k & 14 & 96.2 & 2.8 & 82810 & 5098 & 4855 \\ \hline
\end{tabularx}
\label{tab:throughput-latency}
\end{table*}

\section{Results}\label{sec:results}

The performance of the NorthPole LLM inference system was evaluated using the open-source IBM Granite-3.3-8b-instruct model\cite{granite-3.3-8b-instruct}, which has 8 billion parameters and is trained for a variety of tasks including general language, mathematics, coding, and instruction following.

\subsection{Model Quantization \& Accuracy}

The original Granite-3.3-8b-instruct model was trained with bfloat16 precision. To evaluate on NorthPole, the trained model was quantized to use $8$-bit activations, 8-bit caches, and 4-bit weights (A8-C8-W4) and accuracy was recovered by applying the SiLQ quantization-aware-training algorithm\cite{esser2025silqsimplelargelanguage} to fine-tune the quantized model. Fine tuning was performed on a single node consisting of 8~H100 GPUs using a batch size of 128 and sequence length of 1024. The quantized model converged within 2 weeks and 128,000 steps.

Accuracy was evaluated on 19 benchmarks comprising common sense reasoning tasks and other tasks taken from versions 1 and 2 of the Open LLM Leaderboard\cite{open-llm-leaderboard}. Averaging across all metrics, the quantized model running on NorthPole matches the accuracy of the original model (56.8 and 56.4, respectively). Individual benchmark scores are shown in \autoref{fig:accuracy}.

\subsection{System Performance}
Latency and throughput measurements were performed on the NorthPole system by running the Granite-3.3-8b-instruct model for 2k and 4k context lengths, where prompt-prefill and token-generation stages were fixed to half of the total context in order to ensure results reflected an equal weighting to each of the phases. For each experiment, 1400 requests were issued and both latency and throughput sequence metrics were collected.
\footnotetext{See text for metric definitions.}

\vspace{1em}
\noindent For each sequence \(s\) or batch \(B\):
\begin{itemize}
    \item \(N^{\text{in}}_{\{s,B\}}\): number of input tokens in prompt
    \item \(N^{\text{out}}_{\{s,B\}}\): number of output tokens generated
    \item \(t^{\text{start}}_{\{s,B\}}\): timestamp when prompt prefill begins
    \item \(t^{\text{first}}_{\{s,B\}}\): timestamp when the first output token is obtained
    \item \(t^{\text{end}}_{\{s,B\}}\): timestamp when sequence generation completes
    \item \(t^{(k)}_{\{s,B\}}\): timestamp when the \(k\)-th output token is obtained (\(k=1,\dots,N^{\text{out}}_{\{s,B\}}\)), so \(t^{(1)}_{\{s,B\}} = t^{\text{first}}_{\{s,B\}}\)
\end{itemize}
\vspace{1em}

\subsubsection{Latency}
Latency is measured with the following metrics:
\vspace{1em}
\begin{itemize}
    \item \textbf{Time-to-First-Token:} The time to produce the first generated token for a sequence from the start of prompt-prefill. \[
        \mathrm{TTFT}_s = t^{\text{first}}_s - t^{\text{start}}_s.
    \]
    \item \textbf{Inter-Token-Latency:} The average time between generated tokens for a given sequence. \[
    \mathrm{ITL}_s = \frac{1}{N^{\text{out}}_s - 1}\sum_{k=2}^{N^{\text{out}}_s}\!\big(t^{(k)}_s - t^{(k-1)}_s\big), \; N^{\text{out}}_s \ge 2
\]
\end{itemize}
\vspace{1em}

The $\mathrm{TTFT}_s$ depends on batch size and the number of input tokens in the sequence, which scale linearly with the number of simultaneous users and the prompt length, respectively. Sequences with 128 tokens (\(N^{\text{in}}_s=64\)) complete prefill within ~5.4~ms on average and those with 4096  (\(N^{\text{in}}_s=2048\)) complete within ~96~ms.

The inter-token latency (per-sequence) is constant across total sequence length, with some variability arising from system dynamics (CPU scheduling, bus arbitration, dynamic batching, etc.). Overall, the $\mathrm{ITL}_s$ $\approx2.8$~ms for a given request.
\vspace{1em}

\subsubsection{Throughput}

Throughput is measured with the following metrics:
\vspace{1em}
\begin{itemize}
    \item \textbf{Input-Tokens-per-Second:} The number of input tokens in a batch, divided by the duration of prompt prefill for that batch. \[
    \mathrm{ITPS}_B = \frac{N^{\text{in}}_B}{\mathrm{TTFT}_B}
\]
    \item \textbf{Output-Tokens-per-Second:} The number of output tokens in a batch, divided by the time to generate them, excluding prompt prefill. \[
    \mathrm{OTPS}_B = \frac{N^{\text{out}}_B}{t^{\text{end}}_B - t^{\text{first}}_B}
\]
    \item \textbf{Effective-Output-Tokens-per-Second:} The number of output tokens in a batch, divided by the time to generate them, including prompt prefill.\[
    \mathrm{EOTPS}_B = \frac{N^{\text{out}}_B}{t^{\text{end}}_B - t^{\text{start}}_B}
\]
\end{itemize}
\vspace{1em}

The results for 2k and 4k context lengths, and the corresponding batch sizes, are shown in \autoref{tab:throughput-latency}.

For deployments of Granite-3.3-8b-instruct (A8-C8-W4, 6 server nodes per instance), each NorthPole LLM inference rack supports up to 3 instances, delivering output throughput of up to 30,000~tokens/second when fully utilized by 28 simultaneous users, each with 2k context length. With 4k context length, the number of simultaneous users becomes 14 (up to 15,000~tokens/second) due to the tradeoff between context length and the number of simultaneous users whose KV caches must share the available on-chip memory.

\subsection{Rack \& System Power}

For data center deployments, rack-level power requirements were first established by analyzing the baseline power consumption of the servers and then incorporating the power demands of the NorthPole cards within a defined maximum power envelope. The NorthPole cards are configurable, allowing trade-offs between performance and power consumption. These estimates were subsequently validated through direct measurement of both system and NorthPole power consumption under idle and representative load conditions.

For each server, the configured Gigabyte systems exhibited an average idle power consumption of 615~W. Each of the NorthPole cards within the system were allocated a 50~W power envelope, yielding a total of 800~W for the aggregated NorthPole card power (16\ensuremath{\times}50~W). An additional 350~W was reserved for fan-based cooling. To account for power delivery inefficiencies and thermal losses, a 20\% margin was also included. The resulting per-server power envelope was therefore estimated at 2.2~kW, corresponding to approximately 39.6~kW for a rack comprising 18~NorthPole LLM nodes.

Measurements to validate the established power requirements were conducted while the NorthPole system executed the Granite-3.3-8b-instruct workload using a representative input set. The maximum observed card power was applied to each system to confirm the accuracy of the power assumptions. In practice, a 6 system, 84-card Granite-3.3-8b-instruct model deployment consumed 10.0~kW, representing 76\% of the allocated power budget. Extrapolating from this measurement, a three-instance configuration would consume approximately 30~kW of total rack power.

For redundancy, the rack power configuration allows limited overcommitment, reserving approximately 5–10~kW of the provisioned capacity to support a small number of system failovers. This approach avoids the need for full duplication of the power supply while maintaining operational resilience.

\subsection{End-to-End Applications}
\begin{figure}[tp]
    \centering
    \includegraphics[width=.9\linewidth]{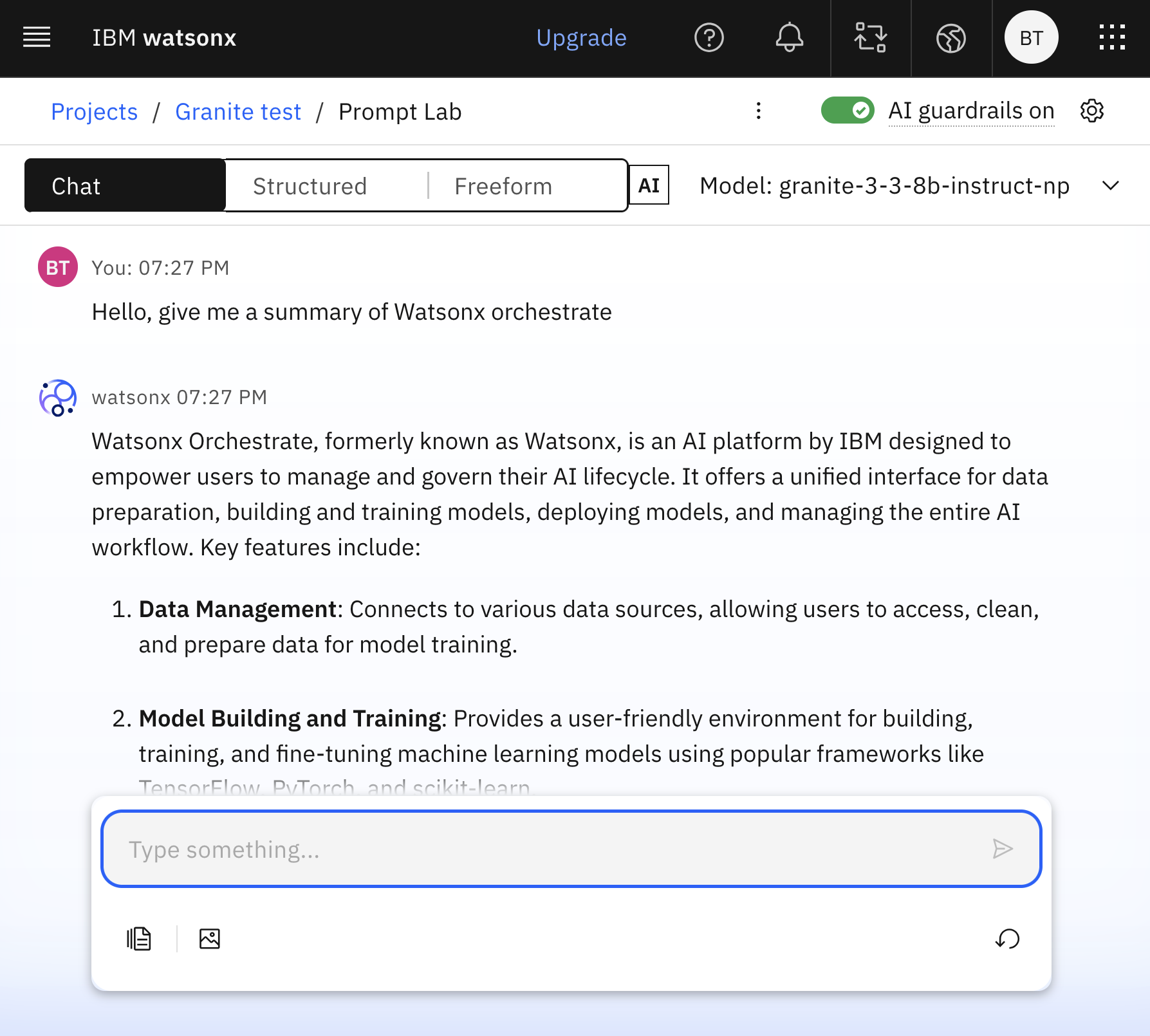}
    \caption{A screenshot of IBM watsonx's chat interface using the IBM NorthPole LLM inference system to run an 8-billion-parameter Granite-3.3-8b-instruct (A8-C8-W4) model.}
    \label{fig:watsonx}
\end{figure}

The cloud-based inference service has been directly integrated with IBM watsonx Orchestrate\cite{IBMwatsonxOrchestrate}, IBM’s unified platform to build, deploy, and manage AI models and workflows. This integration enables a variety of applications which leverage watsonx APIs to use the NorthPole LLM inference systems seamlessly. \autoref{fig:watsonx} demonstrates IBM watsonx prompt lab using Granite-3.3-8b-instruct-np. 

\section{Conclusions \& Next Steps}

Enterprise AI applications are increasingly focused on small language models (SLMs, $<$10~billion parameters), whose rapidly improving capabilities already approach the performance of giant frontier models ($>$100~billion parameters) on targeted benchmarks at a fraction of the cost\cite{belcak2025smalllanguagemodelsfuture}\!\cite{pareja2024unveilingsecretrecipeguide}. The NorthPole LLM inference system is a data-center-ready hardware framework that is particularly suitable for deploying enterprise AI workflows that use SLM agents for low-latency, low-hallucination, low-cost inference. In particular, the architecture of the recently released IBM Granite-4.0 model family (350~million to 32~billion parameters) neatly complements the all-on-chip memory model of the NorthPole accelerator\cite{granite-4.0, granite-4.0-nano}. Continuing to serve the Granite series of models, and their evolutions, on the NorthPole LLM inference system will provide many opportunities to further exploit the vertical integration of the co-designed hardware-software LLM inference stack.

\section*{Acknowledgment}
This work builds on previously published research\cite{np_science} that was supported by the United States Air Force under Contract No. FA8750-19-C-1518.
The authors are grateful to Deepika Bablani for technical contributions.

%% file: references.bib
@article{np_science,
  title={Neural inference at the frontier of energy, space, and time},
  author={Modha, Dharmendra S and Akopyan, Filipp and Andreopoulos, Alexander and Appuswamy, Rathinakumar and Arthur, John V and Cassidy, Andrew S and Datta, Pallab and DeBole, Michael V and Esser, Steven K and Otero, Carlos Ortega and others},
  journal={Science},
  volume={382},
  number={6668},
  pages={329--335},
  year={2023},
  publisher={American Association for the Advancement of Science}
}

@inproceedings{np_isscc,
  title={{IBM NorthPole}: An Architecture for Neural Network Inference with a 12nm Chip},
  author={Cassidy, Andrew S and Arthur, John V and Akopyan, Filipp and Andreopoulos, Alexander and Appuswamy, Rathinakumar and Datta, Pallab and Debole, Michael V and Esser, Steven K and Otero, Carlos Ortega and Sawada, Jun and others},
  booktitle={2024 IEEE International Solid-State Circuits Conference (ISSCC)},
  volume={67},
  pages={214--215},
  year={2024},
  organization={IEEE}
}

@inproceedings{np_hpec_llm,
  author={Appuswamy, Rathinakumar and Debole, Michael V. and Taba, Brian and Esser, Steven K. and Cassidy, Andrew S. and Amir, Arnon and Andreopoulos, Alexander and Bablani, Deepika and Datta, Pallab and Kusnitz, Jeffrey A. and others},
  booktitle={2024 IEEE High Performance Extreme Computing Conference (HPEC)}, 
  title={Breakthrough Low-Latency, High-Energy-Efficiency {LLM} Inference Performance Using {NorthPole}}, 
  year={2024},
  volume={},
  number={},
  pages={1-8},
  doi={10.1109/HPEC62836.2024.10938418}
}

@online{gigabyte_g292_2g0,
  title     = {{G292-2G0 HPC/AI Server}},
  author    = {{GIGABYTE, Inc.}},
  year      = 2025,
  url       = {https://www.gigabyte.com/Enterprise/GPU-Server/G292-2G0-rev-100},
  urldate   = {2025-10-09}
}

@online{granite-3.3-8b-instruct,
  title     = {{granite-3.3-8b-instruct}},
  author    = {{IBM Corporation}},
  year      = 2025,
  url       = {https://huggingface.co/ibm-granite/granite-3.3-8b-instruct},
  urldate   = {2025-10-15}
}

@online{granite-8b-code-base-4k,
  title     = {{granite-8b-code-base-4k}},
  author    = {{IBM Corporation}},
  year      = 2025,
  url       = {https://huggingface.co/ibm-granite/granite-8b-code-base-4k},
  urldate   = {2025-10-15}
}

@online{open-llm-leaderboard,
  title     = {{Open LLM Leaderboard}},
  author    = {{Hugging Face, Inc.}},
  year      = 2025,
  url       = {https://huggingface.co/open-llm-leaderboard},
  urldate   = {2025-10-15}
}

@online{granite-4.0,
  title     = {{IBM Granite 4.0}: hyper-efficient, high performance hybrid models for enterprise},
  author    = {Kate Soule and Dave Bergmann},
  year      = 2025,
  url       = {https://www.ibm.com/new/announcements/ibm-granite-4-0-hyper-efficient-high-performance-hybrid-models},
  urldate   = {2025-10-31}
}

@online{granite-4.0-nano,
  title     = {{Granite 4.0 Nano}: Just how small can you go?},
  author    = {Kate Soule and Rameswar Panda},
  year      = 2025,
  url       = {https://huggingface.co/blog/ibm-granite/granite-4-nano},
  urldate   = {2025-10-31}
}

@online{IBMwatsonxOrchestrate,
  title     = {{watsonx Orchestrate}},
  author    = {{IBM Corporation}},
  year      = 2025,
  url       = {https://www.ibm.com/products/watsonx-orchestrate},
  urldate   = {2025-10-31}
}

@misc{huang2019gpipeefficienttraininggiant,
      title={{GPipe}: Efficient Training of Giant Neural Networks using Pipeline Parallelism}, 
      author={Yanping Huang and Youlong Cheng and Ankur Bapna and Orhan Firat and Mia Xu Chen and Dehao Chen and HyoukJoong Lee and Jiquan Ngiam and Quoc V. Le and Yonghui Wu and Zhifeng Chen},
      year={2019},
      eprint={1811.06965},
      archivePrefix={arXiv},
      primaryClass={cs.CV},
      url={https://arxiv.org/abs/1811.06965}, 
}

@misc{shazeer2018meshtensorflowdeeplearningsupercomputers,
      title={{Mesh-TensorFlow}: Deep Learning for Supercomputers}, 
      author={Noam Shazeer and Youlong Cheng and Niki Parmar and Dustin Tran and Ashish Vaswani and Penporn Koanantakool and Peter Hawkins and HyoukJoong Lee and Mingsheng Hong and Cliff Young and Ryan Sepassi and Blake Hechtman},
      year={2018},
      eprint={1811.02084},
      archivePrefix={arXiv},
      primaryClass={cs.LG},
      url={https://arxiv.org/abs/1811.02084}, 
}

@misc{shoeybi2020megatronlmtrainingmultibillionparameter,
      title={{Megatron-LM}: Training Multi-Billion Parameter Language Models Using Model Parallelism}, 
      author={Mohammad Shoeybi and Mostofa Patwary and Raul Puri and Patrick LeGresley and Jared Casper and Bryan Catanzaro},
      year={2020},
      eprint={1909.08053},
      archivePrefix={arXiv},
      primaryClass={cs.CL},
      url={https://arxiv.org/abs/1909.08053}, 
}

@misc{openai2025gptoss120bgptoss20bmodel,
      title={gpt-oss-120b \& gpt-oss-20b Model Card}, 
      author={Sandhini Agarwal and Lama Ahmad and Jason Ai and Sam Altman and Andy Applebaum and Edwin Arbus and Rahul K. Arora and Yu Bai and Bowen Baker and Haiming Bao and others},
      year={2025},
      eprint={2508.10925},
      archivePrefix={arXiv},
      primaryClass={cs.CL},
      url={https://arxiv.org/abs/2508.10925}, 
}

@online{openai_api,
  title     = {{OpenAI API reference}},
  author    = {{OpenAI, Inc.}},
  year      = 2025,
  url       = {https://platform.openai.com/docs/api-reference/introduction},
  urldate   = {2025-10-09}
}

@misc{esser2025silqsimplelargelanguage,
      title={{SiLQ}: Simple Large Language Model Quantization-Aware Training}, 
      author={Steven K. Esser and Jeffrey L. McKinstry and Deepika Bablani and Rathinakumar Appuswamy and Dharmendra S. Modha},
      year={2025},
      eprint={2507.16933},
      archivePrefix={arXiv},
      primaryClass={cs.LG},
      url={https://arxiv.org/abs/2507.16933}, 
}

@misc{belcak2025smalllanguagemodelsfuture,
      title={Small Language Models are the Future of Agentic {AI}}, 
      author={Peter Belcak and Greg Heinrich and Shizhe Diao and Yonggan Fu and Xin Dong and Saurav Muralidharan and Yingyan Celine Lin and Pavlo Molchanov},
      year={2025},
      eprint={2506.02153},
      archivePrefix={arXiv},
      primaryClass={cs.AI},
      url={https://arxiv.org/abs/2506.02153}, 
}

@misc{pareja2024unveilingsecretrecipeguide,
      title={Unveiling the Secret Recipe: A Guide For Supervised Fine-Tuning Small {LLMs}}, 
      author={Aldo Pareja and Nikhil Shivakumar Nayak and Hao Wang and Krishnateja Killamsetty and Shivchander Sudalairaj and Wenlong Zhao and Seungwook Han and Abhishek Bhandwaldar and Guangxuan Xu and Kai Xu and others},
      year={2024},
      eprint={2412.13337},
      archivePrefix={arXiv},
      primaryClass={cs.LG},
      url={https://arxiv.org/abs/2412.13337}, 
}

@article{economist-2025-llm-waning,
  author  = {{The Economist}},
  title   = {Faith in God-like large language models is waning},
  journal = {The Economist},
  year    = {2025},
  month   = sep,
  day     = {8},
  url     = {https://www.economist.com/business/2025/09/08/faith-in-god-like-large-language-models-is-waning}
}

@article{wsg-2025-small-models,
  author  = {Mims, Christopher},
  title   = {Large Language Models Get All the Hype, but Small Models Do the Real Work},
  journal = {The Wall Street Journal},
  year    = {2025},
  month   = oct,
  day     = {31},
  url     = {https://www.wsj.com/tech/ai/large-language-models-get-all-the-hype-but-small-models-do-the-real-work-225d3145}
}

@article{Noffsinger2025,
  author       = {Noffsinger, Jesse and Patel, Mark and Sachdeva, Pankaj and Bhan, Arjita and Chang, Haley and Goodpaster, Maria},
  title        = {The cost of compute: A \$7 trillion race to scale data centers},
  journal      = {McKinsey \& Company Insights},
  year         = {2025},
  month        = {April},
  day          = {28},
  url          = {https://www.mckinsey.com/industries/technology-media-and-telecommunications/our-insights/the-cost-of-compute-a-7-trillion-dollar-race-to-scale-data-centers}
}

@techreport{Pilz2025,
  author       = {Pilz, Konstantin F. and Mahmood, Yusuf and Heim, Lennart},
  title        = {AI's Power Requirements Under Exponential Growth: Extrapolating AI Data Center Power Demand and Assessing Its Potential Impact on U.S. Competitiveness},
  institution  = {RAND Corporation},
  number       = {RR-A3572-1},
  year         = {2025},
  url          = {https://www.rand.org/content/dam/rand/pubs/research_reports/RRA3500/RRA3572-1/RAND_RRA3572-1.pdf#:~:text=Given%20recent%20training%20compute%20growth%2C,the%20power%20capacity%20needed%20for},
  doi          = {10.7249/RRA3572-1}
}
